\documentclass[aps,pra,twocolumn,showpacs,amsmath,amssymb,groupedaddress]{revtex4-1}
\usepackage{graphicx}
\begin{document}
\title{Theoretical Exploration of Competing Phases of Lattice Bose Gases in a Cavity}
\author{Renyuan Liao}
\author{Huang-Jie Chen}
\author{Dong-Chen Zheng}
\author{Zhi-Gao Huang}
\affiliation{Fujian Provincial Key Laboratory for Quantum Manipulation and New Energy Materials, College of Physics and Energy, Fujian Normal University, Fuzhou 350117, China}
\affiliation{Fujian Provincial Collaborative Innovation Center for Optoelectronic Semiconductors and Efficient Devices, Xiamen, 361005, China}
\date{\today}
\begin{abstract}
    We consider bosonic atoms loaded into optical lattices with cavity-mediated infinite-range interactions. Competing short- and global-range interactions cultivates a rich phase diagram. With a systematic field-theoretical perspective, we present an $\emph{analytical}$ construction of global ground-state phase diagram. We find that the infinite-range interaction enhances the fluctuation of the number density. In the strong coupling regime, we find four branches of elementary excitations with two being ``partilce-like" and two being ``hole-like", and that the excitation gap becomes soft at the phase boundary between compressible phases and incompressible phases. We derive an effective theory describing compressible superfluid and supersolid states. To complement this perturbative study, we construct a self-consistent mean-field theory and find numerical results consistent with our theoretical analysis. We map out the phase diagram and find that a charge density wave may undergo a structure phase transition to a different charge density wave before it finally enters into the supersolid phase driven by increasing the hopping amplitude.
\end{abstract}
\pacs{67.85.Hj, 05.30.JP, 03.75.Hh, 03.75.Lm}
\maketitle
\section{Introduction}
Ultracold gases in optical lattices are one of the most intriguing systems in which the power of atomic and laser physics can be exploited to explore generic phenomena of solid-sate physics~\cite{BLO08}. They have been proven to be impressively successful in simulating strongly correlated models like the Bose-Hubbard models, which features a quantum phase transition from a superfluid to a Mott insulating phase~\cite{JAK98,GRE02}. Recent years have seen major advances in the exploration of many-body systems in which matter is strongly coupled to light~\cite{CAR13}. In particular, recent experimental realization of competing short- and infinite-range interactions~\cite{ETH16} for bosonic atoms in optical lattices has opened a new avenue for exploring new phases of matter. This is achieved by trapping quantum gases in an optical lattice inside a high finesse optical cavity. The infinity-range interactions  is mediated by a vacuum mode of the cavity and can be independently controlled by tuning the cavity resonance~\cite{ETH10,ETH12,RIT13}.

 Essential physics of this system falls into the category of the extended Hubbard model~\cite{DUT15,SAN16}. In the presence of the
 global-range interactions, novel phases such as charge density wave (CDW) and supersolid (SS) phase emerge, in addition to the conventional superfluid (SF) phase and Mott insulating (MI) phase. Understanding the phase diagram and related phase transition has been the focus of recent studies~\cite{LI13,HAB13,ZHA16,DOG16,NIE16,MUE16,PAN16,FLO17}. These studies mainly concentrated on global phase diagrams and drew heavily on sophisticated numerical methods, making the study of the properties of these phases and analytical methods to understand the underlying physics a useful complement.

 In this study, we shall explore the relevant physics transparently by employing both analytical approach and numerical approach. The findings of our study are two folds: On the one hand, in the atomic limit, we obtain analytically the ground state energy density. With this, we construct the phase diagram consistent with previous numerical approach~\cite{MUE16,FLO17}. Based on this, we carry out a field-theoretical analysis~\cite{FIS89,FRE96,NEL98,STO01,SEN05,SAN09,BRA09}, by which physics close to phase boundaries between compressible phases and non-compressible phases can be qualitatively examined. On the other hand, we construct a a self-consistent local mean-field theory which is numerically cheap. With this, we find interesting structural phase transitions between different charge density wave driven by hopping amplitudes before it enters the supersolid phase.

The paper is structured as follows: In Sec. II the model is introduced. We present a functional integral formulation of this problem. In the atomic limit, we construct the ground state phase diagram analytically. Then we proceed to study physics close to compressible and incompressible boundary by carrying out perturbative expansion on the hopping parameter which is assumed to be small. In Sec. III, we formulate a self-consistent mean-field theory, by which properties of compressible phases are investigated. Finally, in Sec. IV, the conclusions are drawn.

\section{Model and Field-theoretical treatment}
We consider the system described by the following canonical Hamiltonian realized very recently~\cite{ETH16}
\begin{eqnarray}
 \hat{H}&=&-\sum_{<ij>}\left(t_{ij}\hat{b}_i^\dagger \hat{b}_j+h.c.\right)+\frac{U}{2}\sum_i\hat{n}_i(\hat{n}_i-1)\nonumber\\
 & &-\frac{K}{M}\left[\sum_{i\in {e}}\hat{n}_i-\sum_{i\in {o}}\hat{n}_i \right]^2-\sum_i\mu \hat{n}_i.
\end{eqnarray}
Here £¬$t_{ij}$ is the hopping matrix element between site $i$ and site $j$, $\hat{b}_i^\dagger$ and $\hat{b}_j$ are the bosonic operators satisfying commutation relation $[\hat{b}_i^\dagger,\hat{b}_j]=\delta_{ij}$, $\hat{n}_i=\hat{b}_i^\dagger \hat{b}_i$ is the associated number operator which counts the particle number on site $i$, and $\mu$ is the chemical potential. The subscript $e$ ($o$) refers to even (odd) lattice sites $i=(i_x,i_y)$ of square lattice potential defined as $i_x+i_y\in$ even (odd), and $<$$ij$$>$ denotes pair of site $i$ and $j$. The on-site repulsive interaction is characterized by $U$, while the infinite-range attractive interaction is denoted by $K$, which favors overall particle number imbalance between even and odd sites.  Interplay of three energy scales is expected to give rise to a multitude of ground-state manifolds. \\

Within the framework of Euclidean functional integral, the partition function of the system may be cast as $\mathcal{Z}=\int \mathcal{D}[b_i^*,b_i]e^{-S}$ with the action given by~\cite{STO09,SIM10} $S=\int_0^\beta d\tau\sum_i \left[b_i^*\partial_\tau b_i +H(b_i^*,b_i)\right]$, here $\beta=1/k_BT$ is the inverse temperature. To decouple the off-site terms in the action, we introduce a real field $\theta(\tau)$ and  complex bosonic fields $\Psi_i(\tau)$ by performing Hubbard-Stratonovich transformations, resulting in an equivalent representation of the partition function
\begin{eqnarray}
&&\mathcal{Z}=\int \mathcal{D}[\Psi_i^*,\Psi_i]\int\mathcal{D}[\theta,b_i^*,b_i]e^{-S_{R}},
\end{eqnarray}
where the resultant action is given by $S_{R}=\int_0^\beta d\tau \sum_{ij}\Psi_i^*t^{-1}_{ij}\Psi_j+S_0+S_I$ with
\begin{eqnarray}
S_0&=&\int_0^\beta d\tau\sum_i\left[K\theta^2-2K\theta(-1)^{i_x+i_y}b_i^* b_i\right]\nonumber\\
&+&\sum_i \left[b_i^*(\partial_\tau-\mu)b_i+\frac{U}{2}b_i^*b_i^*b_ib_i\right],\\
S_I&=&-\int_0^\beta d\tau \sum_i\left(\Psi_i^*b_i+b_i^*\Psi_i\right).
\end{eqnarray}
Before embarking on detailed analysis with field-theoretical machinery, we make some comments. The free part $S_0$ is readily solvable since the corresponding Hamiltonian $\hat{H}_0(\theta)=\sum_i\hat{H}_{0i}(\theta)$ can be diagonalized in the occupation number representation. With the interacting part $S_I$ present, the physics could not be solved in a close form, however, we can gain physical insights by seeking perturbative expansion on top of $S_0$ in terms of fields $\Psi_i$, which serves as the superfluid order parameter.

Now we subject the action $S_{R}$ to a saddle point analysis. Extremum of variation of the action with respect to $\theta(\tau)$ yields $\theta=\sum_i(-1)^{i_x+i_y}<\!\hat{n}_i\!>/M$. The physical meaning is clear: $M\theta$ counts the particle number difference between even sites and odd sites, and $\theta$ could be regarded as an order parameter representing charge degrees of freedom.

\begin{figure}[tb]
\includegraphics[width=1.0\columnwidth,height=1.0\columnwidth]{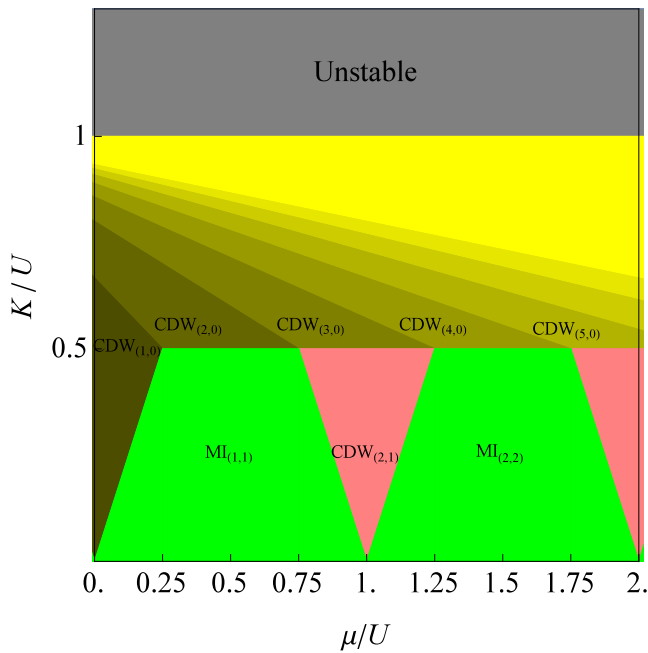}
\caption{(color online). Global ground-state phase diagram spanned by $\mu/U$ and $K/U$ at atomic limit ($t_{ij}/U=0$). The phase diagram can be loosely divided into three regimes depending on the strength of infinite-range interaction: (1)$K/U\in(0,0.5)$, the system is either in a Mott insulating (MI) phase with $n_e=n_o$ or in a partially polarized charge density wave (CDW) phase with $n_e-n_o=1$, where we always assume $n_e\ge n_o$ as the system enjoys an Ising-type $\mathbb{Z}_2$ symmetry; (2)$K/U\in(0.5,1)$, the system is in a fully polarized CDW phase with $n_o=0$; (3)$K/U>1$, the system is unstable toward collpase.}
\label{fig1}
\end{figure}

\begin{figure}[tb]
\includegraphics[width=1.0\columnwidth]{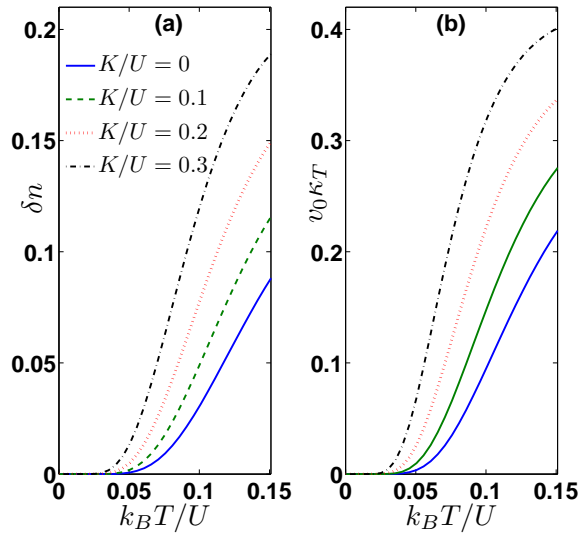}
\caption{(color online). Temperature dependence of (a) the variation of particle number $\delta n$ on one supercell (with one even site and one odd site) and (b) isothermal compressibility $\kappa_T$ for $MI_{(1,1)}$ phase at different infinite-range interaction strength $K/U=0$, $0.1$, $0.2$ and $0.3$. Here $\mu/U=0.60$, and $v_0$ is the volume of one supercell.}
\label{fig2}
\end{figure}

Let's consider the atomic limit where the hopping amplitude between sites is negligible ($t_{ij}/U=0$), and the resultant action reduces to a free one: $S_R=S_0$. The eigenvalue corresponding to $\hat{H}_0$ for per ``supercell" (with one oven and one odd sites) after taking account of the self-consistency conditions for $\theta$ is given by
\begin{eqnarray}
  &&E(n_e,n_o)=\frac{U}{4}\left[(n_e+n_o)-(1+\frac{2\mu}{U})\right]^2\nonumber\\
  &&+\frac{U}{4}\left[(1-\frac{2K}{U})(n_e-n_o)^2-(1+\frac{2\mu}{U})^2\right].
\end{eqnarray}
Here $n_e(n_o)$ represents the occupation number of one even (odd) site. The ground state is achieved by minimizing the eigenvalue $E(n_o,n_e)$ with respect to quantum numbers $n_e$ and $n_o$. Since the system enjoys an Ising-type $\mathbb{Z}_2$ symmetry corresponding to exchange of even and odd sites, we may choose $n_e\ge n_o$ from now on. To facilitate the analysis, we define $1+2\mu/U=n+x$, with $n=int\left[1+2\mu/U\right]$ being the integer closest to $1+2\mu/U$, and $x\in(-1/2,1/2)$. Firstly, let's consider the case $K/U\in(0,1/2)$, the system is in a Mott insulating (MI) phase with $n_e=n_o=n/2$ if $n$ is even. If $n$ is odd, then the system is in the MI phase when $K/U<|x|$, and in a partially polarized charge density wave (CDW) phase with $n_e=n_o+1=(n+1)/2$, vice versa. Secondly, if $K/U\in(1/2,1)$, the system enters into a fully polarized CDW phase with $n_e=int\left[\frac{1+2\mu/U}{2(1-K/U)}\right]$ and $n_o=0$. Finally, if $K/U>1$, the ground energy is unstable toward collapse since it corresponds to an infinite filling. The above discussions for the ground-state phase diagram are summarized in Fig.~$\ref{fig1}$.\\
The low temperature properties of the system may be captured by only considering the particle and hole excitations on one supercell~\cite{GER07}, since tunneling between sites are completely neglected and hence the system consists of isolate pair sites.  For brevity, let us define $C_p=\sum_{s=e,o} e^{-\beta E_{sp}}$ and $C_h=\sum_{s=e,o} e^{-\beta E_{sh}}$, where $E_{sp}$ and $E_{sh}$ is the particle and hole excitation on $s=(e/o)$ site, respectively.  To put it explicitly, $E_{ep}=E(n_e+1,n_o)-E(n_e,n_o)$ and $E_{eh}=E(n_e-1,n_o)-E(n_e,n_o)$, and similarly for $E_{op}$ and $E_{oh}$. The partition function on one supercell is therefore approximated as $z_0=e^{-\beta E(n_e,n_o)}(1+C_p+C_h)$. The variation of density fluctuations on a  supercell is given by $\delta n=(C_p-C_h)/(1+C_p+C_h)$.  Equivalently interesting is the square density fluctuations $\delta n^2\equiv \left<n^2\right>-\left<n\right>^2$, which is related to the isothermal compressibility via thermodynamic relation $\delta n^2=\left<n\right>^2\kappa_T v_0/\beta$ with $v_0$ being the volume of one supercell. We find that $\delta n^2=(C_p+C_h+4C_pC_h)/(1+C_p+C_h)^2$. The temperature dependence of $\delta n$ and $\kappa_T$ is shown in Fig.~\ref{fig2}. At zero temperature, the particle number fluctuation and thermal compressibility is frozen out, indicating its non-compressible nature.  It clearly indicates that the larger $K/U$ is, the larger thermal fluctuation it induces. We attribute this fluctuation-enhancing behavior to the effects of the infinite-range interactions.

We proceed to take into account the effects of a finite hopping amplitude. We can evaluate the partition function by performing Taylor expansion in the exponent
\begin{eqnarray}
 \frac{\mathcal{Z}}{\mathcal{Z}_0}=\int \mathcal{D}[\Psi_i^*,\Psi_i]e^{-\int d\tau\sum_{ij}\Psi_i^*t^{-1}_{ij}\Psi_j}\left<\sum_{l=0}\frac{S_I^l}{l!}\right>_0,
\end{eqnarray}
where $\left<O\right>_0=(\int e^{-S_0}O)/(\int e^{-S_0})$. To the quadratic order in the fields $\Psi_i$, by transforming to momentum-frequency representation, we obtain $\mathcal{Z}/\mathcal{Z}_0=\int \mathcal{D}[\Psi^*(k),\Psi(k)]e^{-S_g}$ with the gaussian action given by
\begin{eqnarray}
  S_g=\frac{M}{2}\sum_{k=(\mathbf{k},iw_n)} \Psi^\dagger(k) \mathcal{G}^{-1}(k)\Psi(k),
\end{eqnarray}
where we have defined $\Psi(k)=(\Psi_e(k),\Psi_o(k))^T$, and used a shorthand notation $k=(\mathbf{k},iw_n)$, with $w_n$ being the bosonic Matsubara frequencies. The inverse Green's function assumes the form of $2\times2$ matrix
\begin{eqnarray}
-\mathcal{G}^{-1}=\begin{pmatrix}
 \frac{n_e+1}{E_{ep}-iw_n}+\frac{n_e}{E_{eh}+iw_n} &-\tilde{t}^{-1}(\mathbf{q})\\ -\tilde{t}^{-1}(\mathbf{q}) &
 \frac{n_o+1}{E_{op}-iw_n}+\frac{n_o}{E_{oh}+iw_n}
\end{pmatrix}.
\end{eqnarray}
 In the above, $\tilde{t}^{-1}(\mathbf{q})$ is the Fourier transform of $t^{-1}_{ij}$. For convenience we shall consider the nearest neighbor hopping only with amplitude $t$, then $\tilde{t}^{-1}(\mathbf{q})=1/[2t\sum_{j=1}^d\cos{(k_j\lambda/2)]}$, with $d$ being the dimension of the system and $\lambda$ being the wavelength of the laser field forming the optical lattices.

\begin{figure}[tb]
\includegraphics[width=1.0\columnwidth]{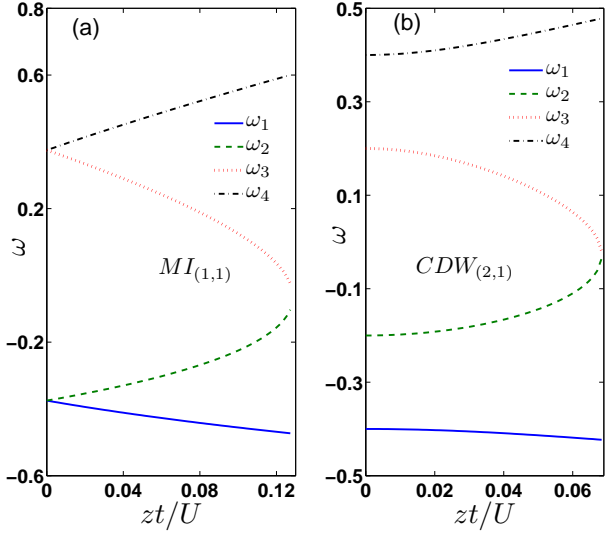}
\caption{(color online). Four branches of excitation spectrum $\omega_i$ ($i$=1..4) at momentum $\mathbf{k}=(\vec{0},0)$ as a function of tunneling parameter $zt/U$ for (a) the $MI_{(1,1)}$ phase with $\mu/U=0.50$ and $K/U=0.25$; (b) the $CDW_{(2,1)}$ phase with $\mu/U=1.0$ and $K/U=0.40$.  The upper two branches are particle excitations and the lower two branches are hole excitations. For both insulating phases, there exists an energy gap for a particle-hole excitation, manifesting their non-compressible nature. Here $z=4$ is the coordination number for square lattices.}
\label{fig3}
\end{figure}

   The excitation spectrum of the system corresponds to the poles of the Green's function. It can be readily found by seeking solutions $\omega$ for the secular equations $\det \mathcal{G}^{-1}(\mathbf{k},\omega)=0$. It features four branches of excitation spectrum $\omega_i$ ($i$=1..4), as shown in Fig.~$\ref{fig3}$ at $k=(\vec{0},0)$ in terms of the tuning parameter $zt/U$ with $z=4$ being the coordination number of square lattices. In absence of hopping ($zt=0$), the incompressible $MI_{(1,1)}$ phase possesses only one type of particle excitations and one type of hole excitations, while incompressible  $CDW_{(2,1)}$ phase carrying charge order possesses two types of particle excitations and two types of hole excitations. At a finite hopping, these two phases both accommodate two branches of particle excitations and two branches of hole excitations. The minimal energy difference between one particle excitation and one hole excitation corresponds to the energy gap for density fluctuations. This excitation gap becomes soft at the phase boundary where phase transition from a non-compressible phase to a compressible phase occurs.

\begin{figure}[tb]
\includegraphics[width=1.0\columnwidth]{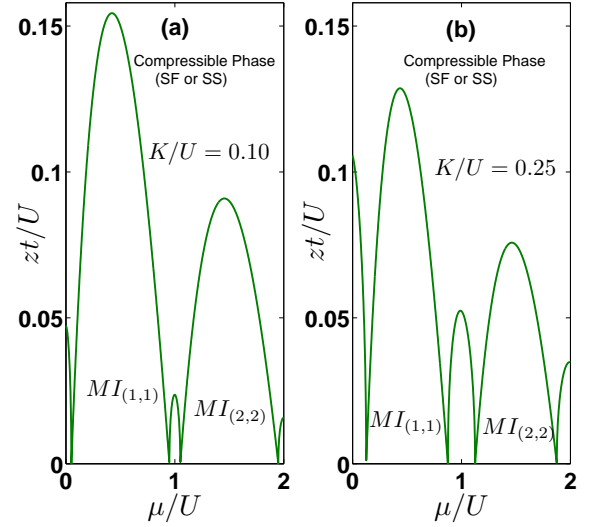}
\caption{(color online). The phase boundary separating compressible (SF or SS) and non-compressible phases (MI or CDW) for two typical infinite-range interaction strengths: (a) $K/U=0.10$  and (b) $K/U=0.25$. Increasing $K/U$ leads to the broadening of the region of $CDW$ phases and the shrinking of the region of $MI$ phases. Here a SF phase stands for a superfluid phase which has off-diagonal long-range order, and a SS phase stands for a supersolid phase which has both diagonal and off-diagonal long-range orders. Here the unlabeled small lobes are  CDW phases generated by the infinite-range interaction.}
\label{fig4}
\end{figure}

The phase boundary separating the superfluid phase and the non-compressible phase occurs~\cite{NEL98,STO01,STO03} at $\det \mathcal{G}^{-1}(0,0)=0$, which yields
\begin{eqnarray}
   \left(\frac{n_e+1}{E_{ep}}+\frac{n_e}{E_{eh}}\right)\left(\frac{n_o+1}{E_{op}}+\frac{n_o}{E_{oh}}\right)=\frac{1}{(zt)^2}.
   \label{Eq:bd}
\end{eqnarray}
 We show the phase boundary in Fig.~$\ref{fig4}$. Evidently the regime of $MI$ phase diminishes as $K/U$ increases. In stark contrast, the regime of $CDW$ phase gets broadened as $K/U$ increases. This suggests that the infinite-range interaction favors the formation of density modulation in the form of a checkerboard pattern with alternating site occupation.\\
To explore the physics of compressible superfluid and supersolid phases, we proceed even further by evaluating the action to the quartic order in order parameter $\Psi_i$, $S=S_0+S_g+S_4$ with $S_4=\sum_{ss^\prime}\sum_{k+l=m+n}u_{ss^\prime}\Psi_s^*(k)\Psi_{s^\prime}^*(l)\Psi_{s^\prime}(m)\Psi_s(n)$, where we may evaluate the coefficients $u_{ss^\prime}$ at zero momentum and zero frequency\cite{SEN05}. By performing derivative expansion, we keep only the most relevant terms in a long-wavelength approximation, and obtain an effective action of Ginzburg-Landau-Wilson type~\cite{HER07, SAC11}
\begin{eqnarray}
  &&S-S_0=\nonumber\\
  &&\int d\tau \frac{d^2\mathbf{r}}{2} \sum_s \left(r_s|\Psi_s|^2+a_s\Psi_s^*\partial_\tau \Psi_s+b_s|\partial_\tau\Psi_s|^2+u_{ss}|\Psi_s|^4\right)\nonumber\\
  &&+\int d\tau \frac{d^2\mathbf{r}}{2}\left[r_{eo}\Psi_e^*(1-\frac{\lambda^2\nabla^2}{16})\Psi_o+c.c+u_{eo}|\Psi_e|^2|\Psi_o|^2\right]\nonumber.\\
\end{eqnarray}
To present the coefficients above in a succinct fashion, we define $A_s=(n_s+1)/E_{sp}^2$ and $B_s=n_s/E_{sh}$. Then the relevant coefficients are given as follows: $r_s=-(A_sE_{sp}+B_sE_{sh})$, $a_s=A_s-B_s$, $b_s=A_s/E_{sp}+B_s/E_{sh}$, $r_{eo}=1/(zt)$, $u_{ss}=(A_s+B_s)(A_sE_{sp}+B_sE_{sh})-A_s(n_s+2)/E_{s2p}-B_s(n_s-1)/E_{s2h}$, and
\begin{eqnarray}
   \frac{u_{eo}}{K}&=&A_eA_o\frac{E_{ep}+E_{op}}{E_{ep}+E_{op}+K}+B_eB_o\frac{E_{eh}+E_{oh}}{E_{eh}+E_{oh}+K}\nonumber\\
   &&-\sum_{s=e,o}A_sB_{-s}\frac{E_{sp}+E_{-sh}}{E_{sp}+E_{-sh}-K}.
\end{eqnarray}
Here $E_{e2p}=E(n_e+2,n_o)-E(n_e,n_o)$ is the ``double particles" excitation energy at even sites and $E_{e2h}=E(n_e-2,n_o)-E(n_e,n_o)$ is the ``double holes" excitation energy at even sites, and similar expressions for $E_{o2p}$ and $E_{o2h}$. The universality class and associated quantum criticality is intimately related to the relevant parameters given above.

 At zero temperature, we assume that field configurations for $\Psi_e$ and $\Psi_o$   are spatially and temporally homogenous. The grand potential $\Omega=- \ln{\mathcal{Z}}/\beta$ of the system reduces to a simple form as follows
\begin{eqnarray}
    \Omega&=&\Omega_0+\sum_{s=e,o}r_s|\Psi_s|^2+r_{eo}(\Psi_e^*\Psi_o+c.c)\nonumber\\
    &&+\sum_{s=e,o}u_{ss}|\Psi_s|^4+u_{eo}|\Psi_o|^2|\Psi_e|^2.
\end{eqnarray}

  Quite generally, the realization of the phase is determined by seeking the global minimum of $\Omega$.  The saddle point condition $\partial \Omega/\partial \Psi_s=0$ yields
  $2r_s\Psi_s+2r_{eo}\Psi_{-s}+4u_{ss}\Psi_s^3+2u_{eo}\Psi_s\Psi_{-s}^2$.
Clearly, if $\Psi_s=0$, then from the above equation we immediately obtain $\Psi_{-s}=0$, namely $\Psi_e$ and $\Psi_o$ vanishes identically at the transition point. The phase boundary is determined by $r_or_e=r_{eo}^2$, which reproduces Eq.~($\ref{Eq:bd}$). Typically close to the phase boundary, the order parameter field satisfies a simple scaling $\Psi_s/\Psi_{-s}=r_{eo}/r_s=\sqrt{r_{-s}/r_s}$. At this level the phase transition is of a continuous one. However, when the system is deep into a superfluid phase with a crystalline order, there may induce a structural transition (where quantum numbers $n_e$ and $n_o$ change) from a CDW phase to another CDW phase. We expect it to be  a first-order one, since it involves a discontinuous change of the free energy. In broken-symmetry phases, the order parameters are determined by the coefficients. The grand potential is fully determined as $\Omega=\Omega_0(n_e,n_o,\mu)+\delta\Omega(n_e,n_o,\mu,zt)$. Given $\mu$ and $zt$, the global minimum of the grand potential is achieved by minimizing over non-negative integer of $n_e$ and $n_o$. However, it should be noted that such perturbative treatment only give qualitatively sensible physics for the regime deep into the compressible phases.
\begin{figure}[tb]
\includegraphics[width=1.0\columnwidth,height=0.6\columnwidth]{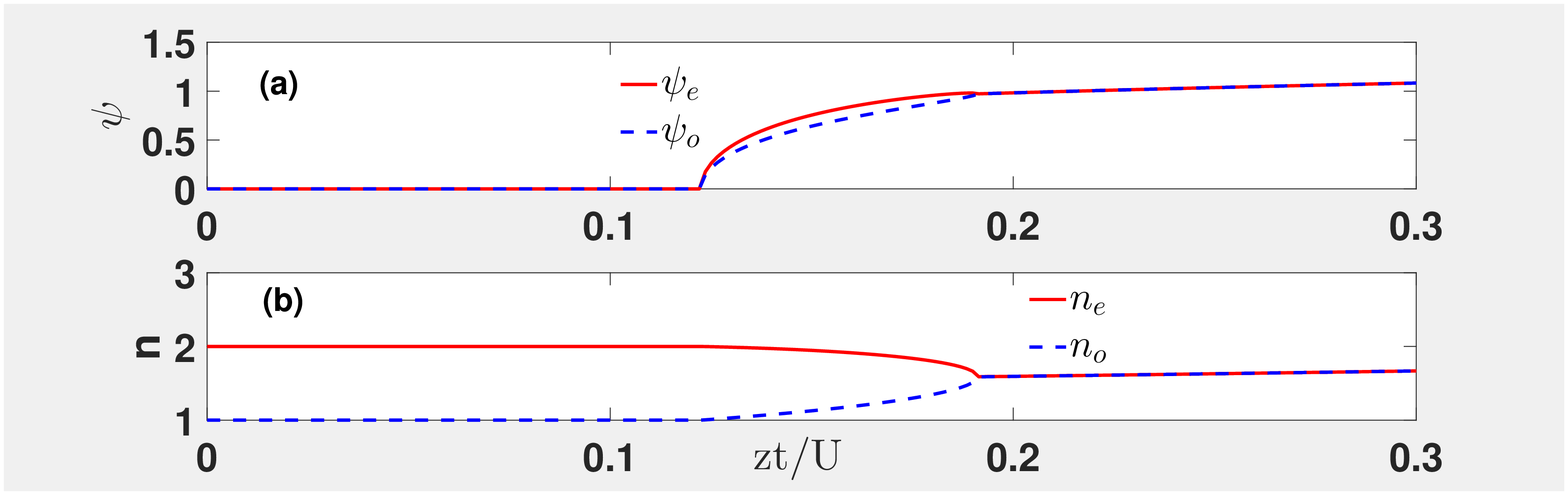}
\caption{(color online). Self-consistent mean-field calculation at zero temperature for $K/U=0.4$ and $\mu/U=1.0$: (a) The magnitude of the order parameters $|\psi_e|$ and $|\psi_o|$ in the supersolid phase as a function of $zt/U$; (b) The density at even site $n_e$ and odd site $n_o$ as a function of $zt/U$. }
\label{fig5}
\end{figure}
\begin{figure}[tb]
\includegraphics[width=1.0\columnwidth,height=0.6\columnwidth]{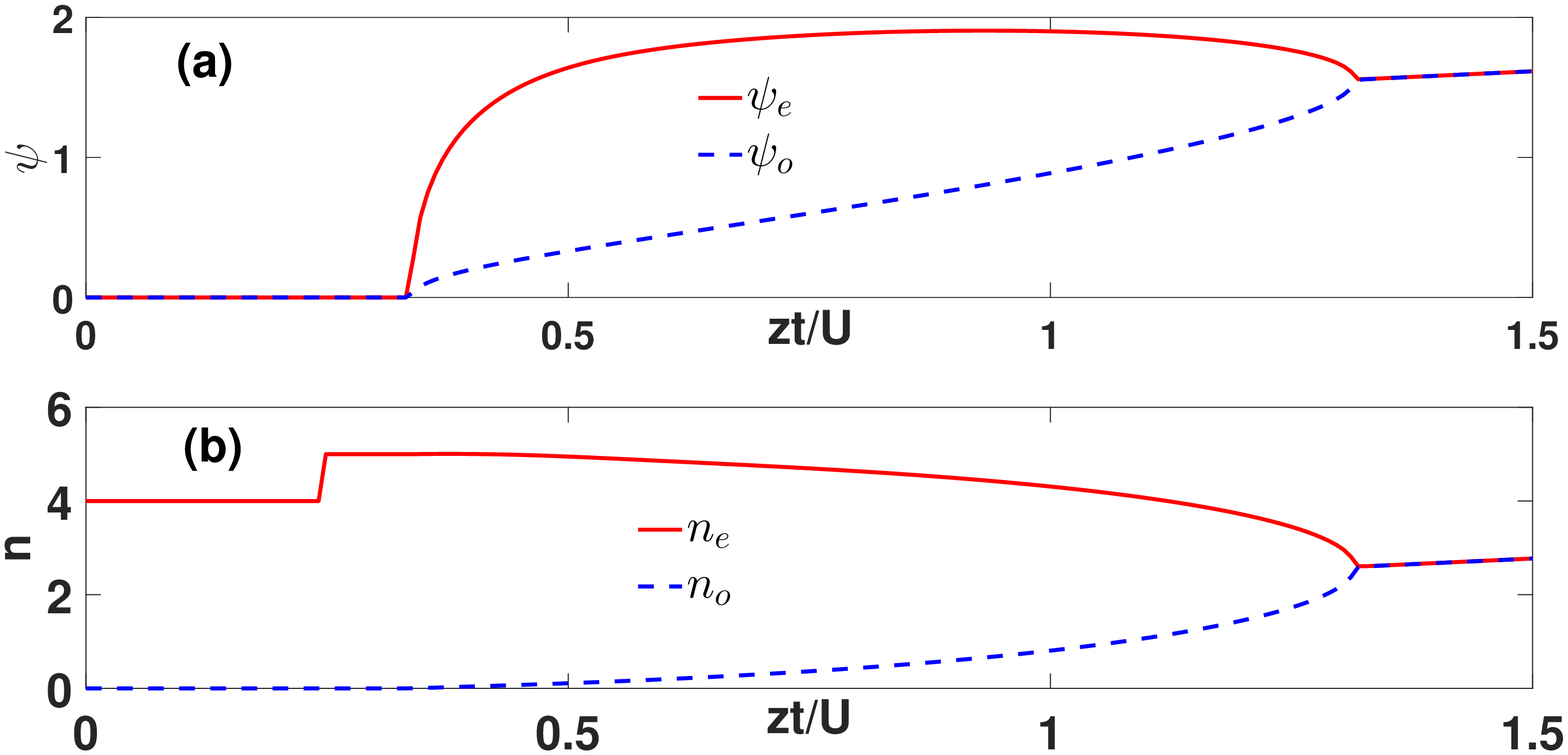}
\caption{(color online). Self-consistent mean-field calculation at zero temperature for $K/U=0.7$ and $\mu/U=1.0$: (a) The magnitude of the order parameters $|\psi_e|$ and $|\psi_o|$ in the supersolid phase as a function of $zt/U$; (b) The density at even site $n_e$ and odd site $n_o$ as a function of $zt/U$. }
\label{fig6}
\end{figure}
 \section{Self-consistent mean-field theory.}
 The perturbative treatment given in the previous section is valid only for a small hopping parameter. To explore physics deep into the compressible phases, we resort to a self-consistent mean-field approximation formulated below.  The mean-field Hamiltonian for a supercell can be constructed as follows:
 \begin{eqnarray}
     &&\hat{H}^{MF}=\sum_{s=e,o}\left[\frac{U}{2}\hat{n}_s(\hat{n}_s-1)-\mu \hat{n}_s\right]-2K\theta(\hat{n}_e-\hat{n}_o)\nonumber\\
     & &-zt\left[\left(\psi_o\hat{b}_e^\dagger+\psi_e^*\hat{b}_o-\psi_o\psi_e^*\right)+h.c.\right]+2K\theta^2.
 \end{eqnarray}

 We may diagonalize $H^{MF}$ in the basis spanned by $|n_e$$>$$\bigotimes$$|n_o$$>$ by simultaneously imposing self-consistency conditions for the charge order parameter $\theta=<\!\hat{n}_e-\hat{n}_o\!>/2$ and for the superfluid order parameters $\psi_e=<\!\hat{b}_e\!>$ and $\psi_o=<\!\hat{b}_o\!>$. It is clear that for self-consistent equations there always exists a trivial solution with $\psi_e=\psi_o$ and $\theta=0$ which corresponds to the SF phase. For a SS phase to be a true ground state, we require that its ground energy is lower than that of a SF one.
  \begin{figure}[tb]
\includegraphics[width=1.0\columnwidth,height=0.6\columnwidth]{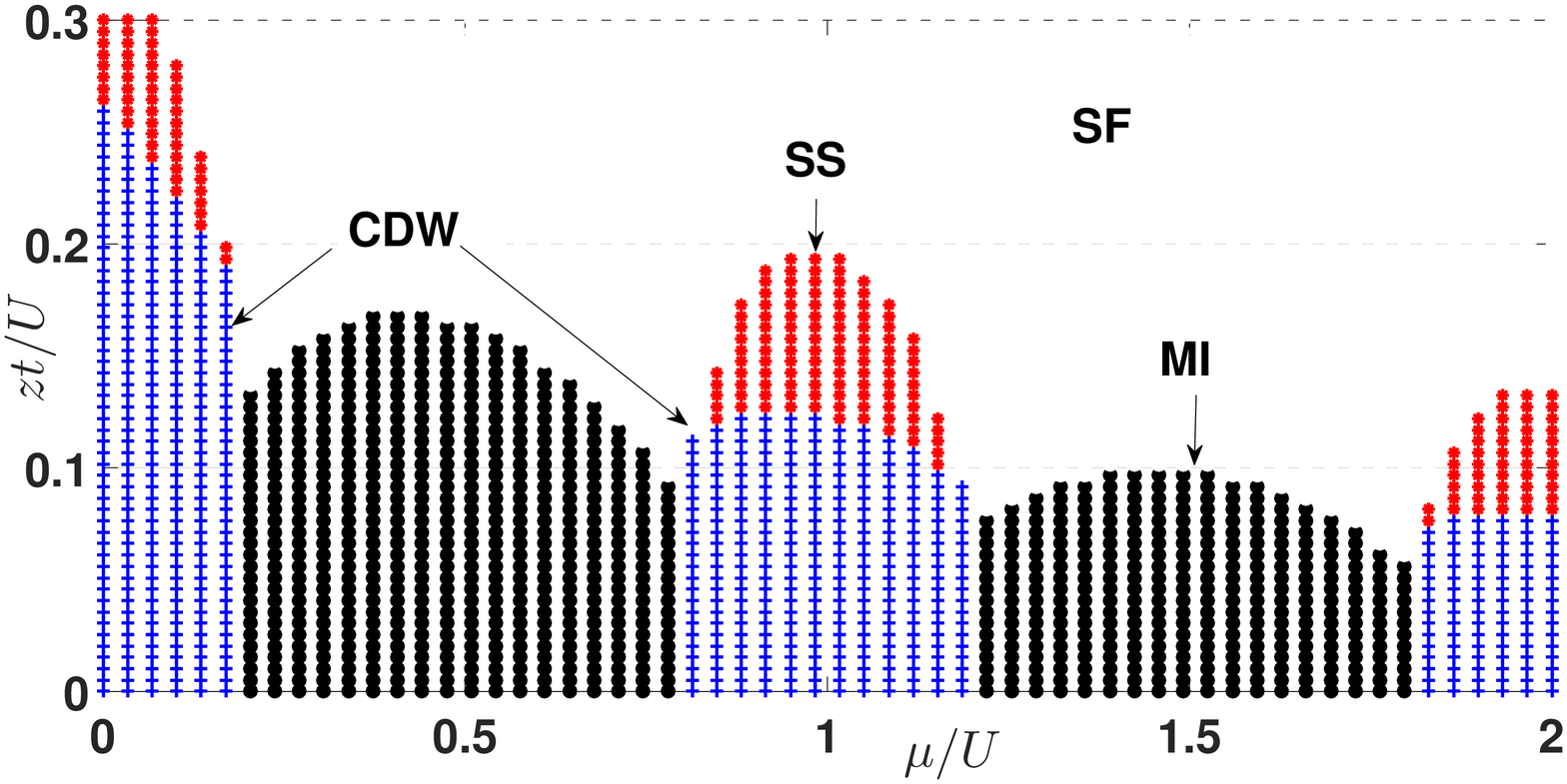}
\caption{(color online). Phase diagram spanned by $zt/U$ and $\mu/U$ from self-consistent mean-field calculation for $K/U=0.4$. }
\label{fig7}
\end{figure}

 The numerical results from this self-consistent theory are shown in Fig.~\ref{fig5}, Fig.~\ref{fig6} and Fig.~\ref{fig7}. For $K/U=0.4$, $\mu/U=1.0$ and $zt/U=0$, the system is evidently in the phase of $CDW_{(2,1)}$, as could be read from Fig.~\ref{fig1}. Now the evolution of the order parameters with respect to the tuning parameter $zt/U$  is shown in Fig.~\ref{fig5}. Across the transition point $zt/U=0.123$, both $\psi_e$ and $\psi_o$ acquires a nonzero value, signaling that the system enters into a SS phase. When $zt/U$ is further increased to $zt/U=0.192$, the system enters into the SF phase with $\psi_e=\psi_o$ and $\phi=0$. This observation is consistent with our general arguments made above based on Landau-type free energy. The behavior of number density clearly follows the steps of the superfluid order parameters. Now let us turn to Fig.~\ref{fig6}, where at $zt/U=0$ the system is in the phase of $CDW_{(4,0)}$ since we choose $K/U=0.7$ and $\mu/U=1.0$ for illustration. As clearly seen in Fig.~\ref{fig6}b, with the increase of hopping parameter $zt/U$, the system first undergoes a structure transition from $CDW_{(4,0)}$ to $CDW_{(5,0)}$, and further increment of $zt/U$ drives  the system to enter a SS phase with  nonzero superfluid order parameter $\psi_e$ and $\psi_o$. With even larger $zt/U$, the system finally favors a SF phase over a SS phase with same superfluid order parameter and particle density at even and odd sites.

 To appreciate how $zt/U$ affects the phase diagram shown in Fig.~\ref{fig1}, we show a phase diagram spanned by $zt/U$ and $\mu/U$  at $K/U=0.4$ in Fig.~\ref{fig7}. Evidently, for a MI phase, increasing of $zt/U$ to some finite value, the system enters into the SF phase; while for a CDW phase, increasing of $zt/U$ first drives the system into a SS phase, and finally into a SF phase with a sufficiently large $zt/U$.
\section{Conclusion}
In summary, we have carried out field-theoretical perturbative study on physics in the strong coupling regime where the hopping parameter is sufficient small. We find that the long-range interaction greatly enhances the thermal fluctuation of the particle number. In the strong coupling regime, we identify four branches of elementary excitation, which corresponds to two types of hole excitation and two types of particle excitation. We derive low-energy effective energy functional for the system in the regime of a small hopping parameter. Finally we construct a self-consistent mean-field theory, by which we find there exists structural phase transition between different CDW phases driven by the hopping parameter. Currently, new research interests~\cite{LEO17,CAS17,ROD17,AND17} including supersolidity breaking translational invariance and relaxation dynamics  are being cultivated along the lines of global collective light-matter interaction.

%

 We acknowledge funding from the NSFC under Grants No. $11674058$, $11274064$ and NCET-$13$-$0734$.

\bibliographystyle{apsrev4-1}
%

\end{document}